\documentstyle[aps,twocolumn,psfig]{revtex}

\begin{document}

\draft
\preprint{}
\title{Surface relaxation and ferromagnetism of Rh\,(001)}
\author{Jun-Hyung Cho and Matthias Scheffler}
\address{Fritz-Haber-Institut der Max-Planck-Gesellschaft, Faradayweg 4-6,
D-14195 Berlin-Dahlem, Germany}
\date{April 28, 1996}

%%%gg
\twocolumn[

\maketitle

%%%gg
\vspace*{-10pt}
%%%gg
\vspace*{-0.7cm}
%%%gg
\begin{quote}
%%%gg
\parbox{16cm}{\small
%%%\begin{abstract}
The significant discrepancy between first-principles calculations and
experimental analyses for the relaxation of the
(001) surface of rhodium has been a puzzle for some years.
In this paper we present 
density functional theory calculations using the local-density
approximation and the generalized gradient approximation of the
exchange-correlation functional. We investigate the thermal expansion of
the surface and the possibility of surface magnetism.
The results throw light on several, hitherto overlooked,
aspects of metal surfaces. We find, that, when the free energy 
is considered, density-functional theory provides results in
good agreement with experiments. \\[0.2cm]
%%%%\end{abstract}
%%%gg
\medskip
%%%%%\pacs{PACS numbers: 68.35.Bs, 75.30.Pd, 68.35.Ja, 63.20.Ry} 
\narrowtext
%%%gg
} 
%%%gg
\end{quote}
%%%gg
]

%%%gg
\narrowtext

The significant discrepancy between first-principles calculations~\cite{Fei90,Met92,Mor93,Cho95,Haf96} and low-energy electron diffraction (LEED)
analyses~\cite{Wat78,Oed88,Beg93} for the relaxation of the (001) surface 
of rhodium has been a puzzle for some years.
The earlier LEED studies~\cite{Wat78,Oed88} concluded
that the interlayer spacing of the surface layer ($d_{12}$) is nearly
identical to that in the bulk ($d_0$), i.e., the top-layer relaxation was
determined to be  ${\Delta}d_{12}/d_0  =  +0.5 \pm 1.0$~\%.
A recent LEED study~\cite{Beg93} found ${\Delta}d_{12}/d_0  = 
-1.16 \pm 1.6$~\%.
On the other hand, first-principles calculations showed a large 
top-layer relaxation ranging from $-$3.2~\% to $-$5.1~\%, depending on
the calculational scheme and/or the employed numerical
accuracy~\cite{Fei90,Met92,Mor93,Cho95,Haf96}.
Inward relaxations are indeed the expected behavior of transition
metals surfaces (see e.g. Ref.~\cite{Met92}
%%%S and references therein
), and the practical zero relaxation determined by LEED is at least
unexpected.

In order to reconcile this disagreement between their calculations and
experiment, Feibelman and Hamann~\cite{Fei90} proposed 
that in the experimental study the metal surface may
be contaminated by
residual hydrogen adsorption
(see also Ref.~\cite{Fei94}). Indeed,
%%%S it is well known that
hydrogen is not easy to detect and
%%%S  that it is
quite soluble in transition metals, such as Ru, Rh, and Pd. 
Furthermore it is known that adsorbed hydrogen significantly reduces
the inward relaxations at metal surfaces
as  it increases the bond coordination of 
the surface atoms, making them, to some extent, more bulk like.
However, the possibility of hydrogen contamination was strongly rejected 
by later experimental papers (e.g.~\cite{Beg93,Men94}).

Morrison {\em et al.}~\cite{Mor93} investigated  an alternative
possibility~\cite{Li91}, namely that the presence of surface magnetism
could increase the first interlayer spacing, i.e., reducing the large inward
relaxation they had obtained in their non-magnetic calculation by ``magnetic
pressure''. In fact,  bulk Rh is already
%%%S  very
close to fulfilling
the Stoner criterion of ferromagnetism, and the narrower density of 
$d$-states at the surface might stabilize a magnetic 
%%%S   ground
state at the surface.
Density-functional theory (DFT) together with the local-density-approximation
(LDA) gives a non-magnetic ground state for Rh\,(001), but
this might be due to the LDA. For example, for bulk iron, which 
is studied in greater detail, the LDA falsely puts the bcc magnetic ground
state at a higher energy than the nonmagnetic hcp and fcc 
states~\cite{harmon,kraft}.
To get around this LDA problem Morrison {\em et al.}~\cite{Mor93}
% created  a Hartree-Fock-like
% pseudopotential and combined it with a spin-polarized DFT-LDA
%calculation for the valence states.
% It is well known that Hartree-Fock (in contrast to the LDA)
% overemphasizes the stability of a magnetic ground state.
employed a pseudopotential which is based upon an atom in which
all the electrons see a Hartree-Fock exchange potential arising from
the core electrons and an LDA potential arising only from the valence
electrons. Then in the surface calculations the valence exchange
potential was taken proportional to $n_{\rm valence}^{1/3}$.
As a consequence, they
found  that their Rh\,(001) surface is ferromagnetic. The magnetic moment is
$M = 1.8~{\mu}_B$/surface atom, and the
resulting magnetic pressure reduced the surface relaxation
${\Delta}d_{12}/d_0$ from  $-3.22$~\% (in the nonmagnetic equilibrium state)
to $-1.52$~\% in the magnetic ground state. Thus these authors concluded
that surface ferromagnetism is the driving force giving rise to the
small surface
relaxation deduced experimentally. Subsequently performed theoretical work, 
however, did not accept their approach and conclusions~\cite{Cho95,Wei93};
and also experimental studies provided no convincing
support~\cite{Wu94}.
In their  spin-polarized photoemission experiment Wu {\em et al.}\cite{Wu94}
found only a
%%%S  very
weak indication of  surface magnetism with a
%%%S   very
small magnetic moment of about $M = 0.2~{\mu}_B$/surface atom.

In this letter we present a new theoretical study which
extends the previous work by considering the generalized
gradient approximation (GGA)~\cite{Per91}, and by taking zero-point
effects and the thermal expansion as well as surface magnetism into account. 
Such a study is desirable since all previous DFT 
calculations~\cite{Fei90,Met92,Mor93,Cho95,Haf96} were performed with the LDA
which does not describe the magnetic state reliably; furthermore,
in all previous work zero-point and thermal vibrations were ignored, while
the LEED data were taken at room temperature~\cite{Wat78,Oed88,Beg93}. 
We will show that
the above noted discrepancy between theoretical and measured results
is mostly due to the unjustified neglect of vibrational contributions
to the free energy. It is argued that the vibrational effects will typically 
play a much bigger role than hitherto anticipated. Furthermore, we find
that
%%%S   although 
surface magnetism
%%%S is indeed present, it is weak and
has a very small effect on the surface interlayer distance.

We employ the full-potential LAPW method~\cite{Bla90,Koh96}
together with norm-conserving pseudopotentials~\cite{Ham89}.
The nonlinear core-valence exchange-correlation
interaction
%%%S ~\cite{Bachelet} 
is treated
%%%S  accurately by 
using the correct core-electron
density as obtained in the atomic calculation~\cite{Cho96a}.
The method gives an accurate and at the same time
computationally efficient description of the interatomic interactions,
total energies, and stable or metastable geometries. Our GGA
calculations are performed consistently by creating  the 
pseudopotential from first-principles DFT-GGA calculations. The
Rh\,(001) surface is modeled by a periodic slab geometry consisting of 
nine layers of Rh and a vacuum thickness corresponding to five such layers.
The geometry is optimized by a damped molecular dynamics~\cite{Koh96}, allowing
the top two layers on both sides of the slab to relax. The remaining atoms
are kept at the
%%%S  positions given by the calculated 
bulk lattice
sites.
%%%S   constant.
The parameters describing the LAPW basis set are:
$({\bf K}^{\rm wf}_{\rm max})^2 = 14$ Ry and
$l^{\rm wf}_{\rm max} = 8$.
For the ${\bf k}$ summation we use 28 points of the 
irreducible part of the surface Brillouin zone.
%%%SFurther details, confirming the high quality of the pseudopotentials 
%%%S and the LAPW basis-set will be given elsewhere.

Since all previous calculations~\cite{Fei90,Met92,Mor93,Cho95,Haf96} 
for Rh\,(001) were performed with the LDA, we also performed LDA calculations, 
which together with our GGA results allow us to examine the
effect of the GGA on the surface properties of Rh\,(001). Using
DFT-LDA our bulk lattice constant is 3.81~\AA{}, which is in good agreement
with  previous calculations~\cite{Met92,Haf96}.
The experimental result, which unlike the quoted
calculated value contains
the influence of zero-point vibrations, is 3.79 \AA~\cite{exp1}.
Were the zero-point vibrations to be included in the theory, the
calculated lattice constant would increase by about 
0.5~\%~\cite{Mor78}.

Using the GGA we find that the bulk lattice constant is expanded with
respect to the LDA value by 2.2~\%, giving it a value of 3.89~\AA.
%%%S It is interesting to note that f
For hcp Ru~\cite{Sta96}
and fcc Pd~\cite{Kol92} a similar trend was found
when comparing LDA and GGA lattice constants (see also Ref. \cite{khein}).
However, we find that the GGA affects the surface relaxation of Rh\,(001)
only
%%%S  very 
little (see Table I), although the cohesive energy,
the surface energy, and the work function are affected noticeably 
compared to the LDA values.

Table I summarizes the results for surface relaxations,
work functions, and  surface energies as obtained
by  different calculations and experiments.
With respect to the surface relaxation it is immediately
evident that the LAPW calculations by Feibelman and Hamann~\cite{Fei90}
give an exceptionally large value. 
The present LDA calculations, those of Cho and Kang~\cite{Cho95},
and those of Methfessel {\em et al.}~\cite{Met92},
who did not relax the second layer, are in good agreement with each other.
Also the result of the nonmagnetic study of Morrison {\em et al.}~\cite{Mor93}
(quoted above) agrees well with our value.
As previously pointed out by Morrison {\em et al.}~\cite{Mor93},
%%%
the too large top-layer relaxation in Feibelman and Hamann's calculations
may be attributed to the use of the poor ${\bf k}$-point 
sampling~\cite{FeiComment}.

The difference between our DFT-LDA results for the
surface relaxation ($\Delta d_{12}/d_0 = -3.0~\%$) 
and the previous LEED analyses~\cite{Wat78,Oed88,Beg93} is decreased significantly compared to the results of Ref. \cite{Fei90,FeiComment} 
(see Table I);
the DFT-GGA calculations give a result
(${\Delta}d_{12}/d_0 = -2.8$~\%) which is even closer.
We will now show that the physics of Rh\,(001) is much
more interesting than previous studies had anticipated.
At first we will address the influence of lattice vibrations of 
the Rh\,(001) surface and show that the restriction to the
$T = 0$~K {\em total energy}\, falsely neglects some important physical aspects, which 
clearly affect the free energy and as a consequence the surface
properties. Then we analyze the possibility of surface magnetism.

\begin{figure}
\psfig{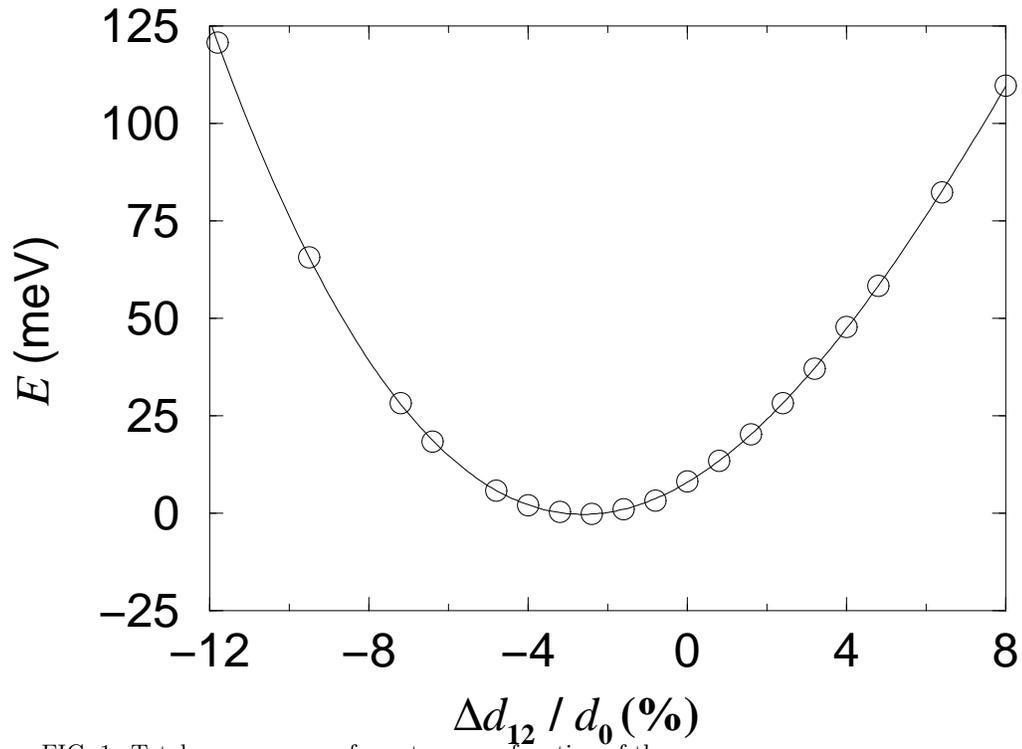}	
\caption{Total energy per surface atom as a function of the top-layer
relaxation for Rh\,(001).
The minimum of the fitted curve is set to be the energy zero.}
\label{V_12}
\end{figure}

\begin{figure}
\psfig{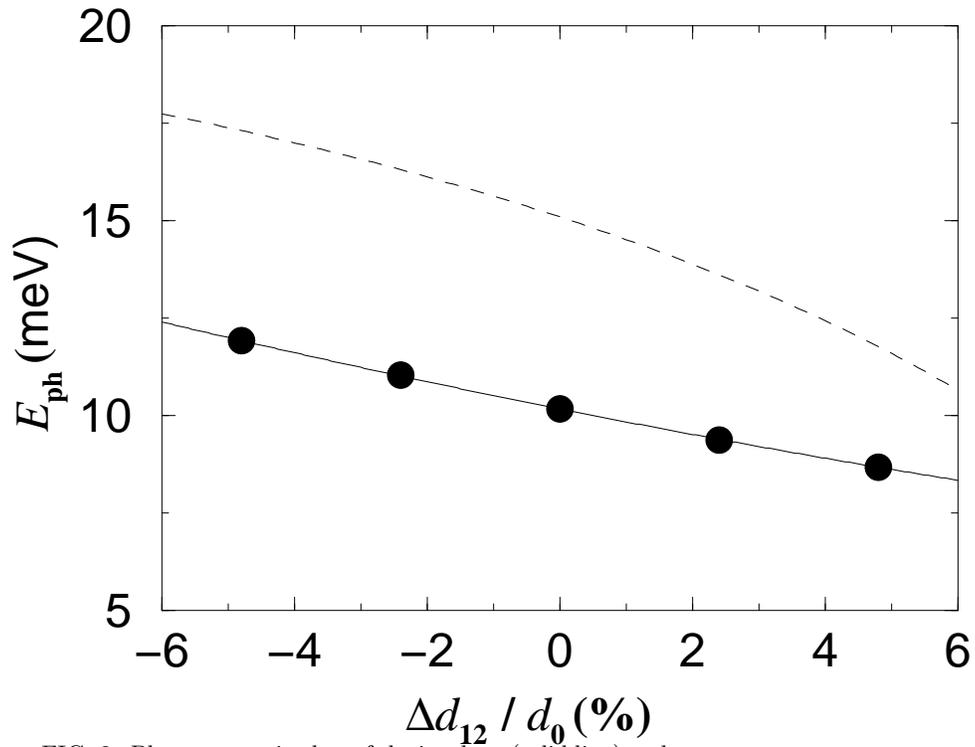}	
\caption{Phonon energies $\hbar \omega_i$ of the in-plane (solid line)
and out-of-plane (dashed line) modes of Rh\,(001)
as a function of the top-layer relaxation.} 
\label{phonon}
\end{figure}

\begin{figure}
\psfig{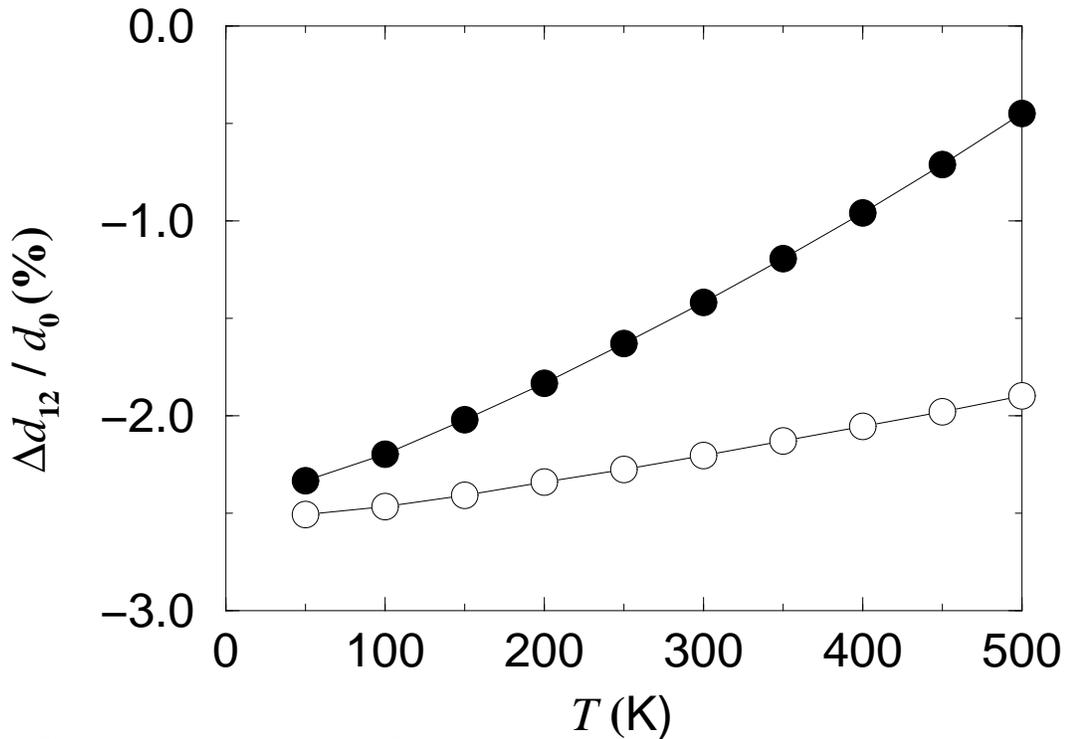}	
\caption{Top-layer relaxation of Rh\,(001) as a function of temperature.
Full dots represent results obtained using eq. (1) with the results 
of Figs. \protect{\ref{V_12}} and \protect{\ref{phonon}}.
Open dots show results obtained if the parallel vibrations 
are neglected.}
\label{d_12(T)}
\end{figure}

It is well known that the zero-point vibrations
give rise to a recognizable effect on the bulk lattice constant.
Moruzzi {\em et al.}~\cite{Mor78} had systematically included this
effect in their KKR studies of metals. Typically, however, 
this effect has been ignored. It is plausible that
vibrational effects may be even larger at surfaces than in the
bulk. In a correct treatment the equilibrium structure at a given temperature
is determined by the minimum of the free energy. At not too high temperatures
this differs from the total energy of the rigid lattice
mainly by the contributions from atomic vibrations to the internal energy (including the zero-point vibrations)
and the vibrational entropy. In the quasi-harmonic approximation the free energy for the surface is $F(T) = Min_{d_{12}} F(d_{12},T)$ with
\begin{eqnarray}
\label{eq:ftot}
F(d_{12}, T)   & = & V(d_{12}) + 
k_B T \sum_{i}
\Bigl\{ {{\hbar \omega_i(d_{12})}\over{2 k_B T}} \nonumber \\
& & + ln \Bigl({ 1 - exp{{- \hbar \omega_i(d_{12})}\over{k_B T}}
}\Bigr)\Bigr\}
\end{eqnarray}
\noindent
where $\hbar \omega_i(d_{12})$ denotes the vibrational frequencies and the sum goes over all bands and ${\bf k}$ points.
The first term in eq. (1) is the first interlayer potential and the second
term is the vibrational energy and entropy.
We note in passing that such a quasi-harmonic description had been used
successfully in DFT calculations of the anomalous thermal expansion
of covalent semiconductors~\cite{Bie89}.
For Ag, Cu, and Al surfaces eq.~(1) has been recently
evaluated by Narasimhan and Scheffler~\cite{Nar96}.
%%%S  who also showed that the surface thermal expansion  is
%%%S  largely due to the vibrations of the top layer
%%%S  parallel to the substrate.
We note that the equilibrium distance $d_{12}$ is shifted away from
the minimum of $V(d_{12})$ towards a larger interlayer spacing
and that this shift is determined by the {\em slope}\, of
the $\hbar \omega_i(d_{12})$ but not their actual values.
To a first approximation this dependence of $\omega_{i}$ on $d_{12}$ 
depends only weakly on the band index and  ${\bf k}$. 
We therefore replaced the sum in eq.~(1) by three 
%%%S appropriate modes, which may be considered as
surface-phonon wave packets.
Only the top layer is moved, and deeper layers are kept fixed.
%%%S Allowing also deeper layers to move changes the results for
%%%S the temperature dependence of the interlayer spacing only little.
Figure~\ref{V_12} provides our DFT-GGA result for the potential energy 
$V(d_{12})$; its  curvature gives the frequencies for the 
perpendicular vibrational mode.
%%%S In order to obtain the frequencies of
For the
the parallel vibrations
we use 
%%%S the frozen-phonon approach, and consider the
two ``modes'' 
along
%%%S  ${\bf x}$ =
$[1 { \overline1} 0]$ and 
%%%S ${\bf y}$ =
$[1 1 0]$, 
which are actually degenerate.
The calculated phonon energies $\hbar \omega_i$
of the in-plane and out of plane vibrations
are shown in Fig.~\ref{phonon}.
Our result for the temperature dependence ${\Delta}d_{12}(T)/d_0$,
considering the three above discussed phonon ``modes'', is given by the full
dots in
Fig.~\ref{d_12(T)}. It is obvious that thermal vibrations
have indeed a pronounced effect. They change the surface relaxation
from the value given by the minimum of the total energy, $-2.8$~\%,
to ${\Delta}d_{12}/d_0 = - 1.4$~\% at 300 K. This result is now
in excellent agreements with that of the room-temperature LEED
analysis~\cite{Beg93} which determined a value of $-1.16 \pm 1.6$~\%.

It is interesting to note that 
%%%S also in this system 
the motion of the
surface layer parallel to the substrate yields the most important 
contributions (compare Ref.~\cite{Nar96}).
If we would neglect the contributions of the parallel motion
and  use only the perpendicular vibration
the resulting top-layer relaxation would be much smaller.
This result,  displayed by the open dots in Fig.~\ref{d_12(T)},
reveals that the anharmonicity of the interlayer potential of
Rh\,(001) does not have a very pronounced influence on the top-layer relaxation.

\begin{figure}
\psfig{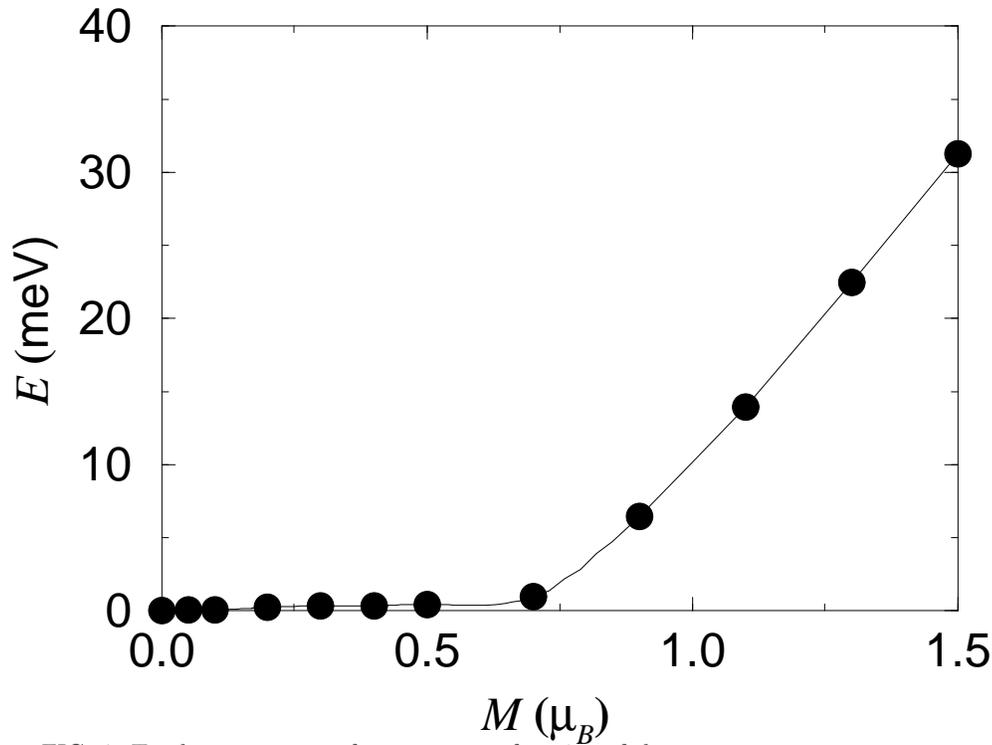}	
\caption{Total energy per surface atom as a function of
the magnetic moment per surface atom for
a surface relaxed by  ${\Delta}d_{12}/d_0 = -2.4$~\%.
The nonmagnetic state defines the energy zero.}
\label{mag.moment}
\end{figure}

Our DFT-GGA calculations predict that the ground state
of Rh\,(001) is nonmagnetic. This result remains even if we 
intentionally increase $d_{12}$ to the unrelaxed geometry, 
thus offering a bigger volume per surface atom which typically helps to 
stabilize a magnetic state.
Despite this apparently clear result of a non-magnetic
ground state, we asked how far away in energy the ferromagnetic
state actually might be.
For this purpose we performed spin-polarized calculations employing the
fixed-spin-moment method~\cite{Wil84}.
Figure~\ref{mag.moment} shows the total energy versus magnetic moment for
a given relaxed surface of ${\Delta}d_{12}/d_0 = -2.4$~\%.
We find that
the total energy monotonically increases with increasing
magnetic moment. This behavior is similar to that of a previous 
fixed-spin-moment study of Cho and Kang~\cite{Cho95},
who used the LDA. 
The present DFT-GGA result for the energy difference $\Delta E$ between 
the nonmagnetic and ferromagnetic states is 
however reduced significantly compared to the previous 
LDA one~\cite{Cho95}, and
Fig.~\ref{mag.moment} shows that $\Delta E$ remains almost constant until the 
magnetic moment reaches a value of 0.5~${\mu}_B$/surface atom\cite{fsm}.
In the fixed-spin-moment method\cite{Wil84}, 
spin-up and spin-down eigenvalues are calculated for different
Fermi energies. For a magnetic moment of $M = 0.5~{\mu}_B$/surface atom
we find that the difference between the two Fermi energies is only
25.9 meV; the total-energy difference at $M = 0.5~{\mu}_B$/surface atom
is only 1 meV.
In other words, our calculations show that the ferromagnetic
state is practically degenerate with the nonmagnetic one,
%%%S . Thus, the system is very close to the critical point,
and we expect that a weak ferromagnetic state will occur 
on Rh\,(001) possibly stabilized by surface imperfections.
This result is consistent with the room temperature spin-polarized
photoemission experiments by Wu {\em et al.}~\cite{Wu94}, 
who observed a rather weak ferromagnetism with the surface magnetic moment
of about 0.2~${\mu}_B$/surface atom.
%%%S  section continued
To some extent our results support the motivation behind the
study of Morrison {\em et al.}~\cite{Mor93}, although
their treatment predicted a rather strong and stable
ferromagnetic state. In contrast to them we find that the magnetic state
is very close to the critical point,
that the magnetic moment should be very small, and thus
the magnetism has practically no effect on the surface relaxation, or
vice versa.

Stimulating discussions with Paul Marcus are gratefully acknowledged.
J.H.C. likes to acknowledge the financial support from
the Korea Science and Engineering Foundation.

\begin{table}
\vspace{1.0cm}
\caption{
Surface relaxations ${\Delta}d_{12}/d_0$ and ${\Delta}d_{23}/d_0$
($d_0$ is the bulk interlayer spacing), 
work functions $\phi$ (eV), and surface energies $\sigma$ (eV/atom)
for Rh\,(001)  as obtained by different calculations and experiments.}
\begin{tabular}{lcccc}
                                   &  ${\Delta}d_{12}/d_0$
                                   &  ${\Delta}d_{23}/d_0$
                                   & $\phi$ 
                                   & $\sigma$\\ \hline
LDA ~\protect{\cite{Fei90}}  &  $-5.1\%$  &  $-0.5\%$  &  5.49  &  1.12\\
LDA ~\protect{\cite{Met92}}  &  $-3.5\%$  &     --   &  5.25  &  1.27\\
LDA ~\protect{\cite{Cho95}}  & $-3.8\%$  &  --  &  --  &  1.29 \\
LDA ~\protect{\cite{Haf96}}  & $-3.8\%$  &  $+0.7\%$  &  --  &  1.44 \\
this -- LDA                        & $-3.0\% $  &  $-0.2\%$ &  5.26  &  1.29\\
this -- GGA                        & $-2.8\% $  &  $-0.1\%$ &  4.92  &  1.04\\
experiments                        & $+0.50 \pm 1.0\%$
\protect{\cite{Oed88}}                        & $ 0 \pm 1.5\%$
\protect{\cite{Oed88}}                                     &  4.65
\protect{\cite{exp1}}                                             &   1.12
\protect{\cite{exp2}}\\
experiments                        & $-1.16 \pm 1.6\%$
\protect{\cite{Beg93}}                        & $ 0 \pm 1.6\%$
\protect{\cite{Beg93}}                                     &  4.98
\protect{\cite{exp3}}                                             &   1.27
\protect{\cite{exp4}}\\
\end{tabular}
\label{geom}
\end{table}

\clearpage

\clearpage

\end{document}